\documentclass[]{spie}  

\usepackage{amsmath,amsfonts,amssymb}
\usepackage{graphicx}
\usepackage[colorlinks=true, allcolors=blue]{hyperref}

\usepackage[final]{microtype}
\usepackage{booktabs}

\title{Harmonization-enriched domain adaptation with light fine-tuning for multiple sclerosis lesion segmentation}

\author[a]{Jinwei~Zhang}
\author[a]{Lianrui~Zuo}
\author[b]{Blake~E.~Dewey}
\author[c,d]{Samuel~W.~Remedios}
\author[a]{\\Savannah~P.~Hays}
\author[e]{Dzung~L.~Pham}
\author[a,c]{Jerry~L.~Prince}
\author[a]{Aaron~Carass}

\affil[a]{Department of Electrical and Computer Engineering, \protect\\ Johns Hopkins University, Baltimore, MD 21218}
\affil[b]{Department of Neurology, Johns Hopkins University, Baltimore, MD 21287}
\affil[c]{Department of Computer Science, Johns Hopkins University, Baltimore, MD 21218}
\affil[d]{Department of Radiology and Imaging Sciences, National~Institutes~of~Health,~Bethesda,~MD~20892,~USA}
\affil[e]{Center for Neuroscience and Regenerative Medicine, Henry M. Jackson Foundation for the Advancement of Military Medicine, Bethesda, MD 20817}

\authorinfo{Further author information: (Send correspondence to Jinwei Zhang)\\Jinwei Zhang: E-mail: jwzhang@jhu.edu}

 \newcommand{\bq}[1]%
{
\vspace*{-0.5em}
\begin{center}%
\begin{minipage}{0.92\columnwidth}%
#1%
\end{minipage}%
\end{center}%
\vspace*{-1em}
}

\begin{document} 
\maketitle

\begin{abstract}
Deep learning algorithms using magnetic resonance (MR) images have demonstrated state-of-the-art performance in the automated segmentation of multiple sclerosis~(MS) lesions.
Despite their success, these algorithms may fail to generalize across sites or scanners, leading to domain generalization errors.
Few-shot or one-shot domain adaptation is an option to reduce the generalization error using limited labeled data from the target domain.
However, this approach may not yield satisfactory performance due to the limited data available for adaptation.
In this paper, we aim to address this issue by integrating one-shot adaptation data with harmonized training data that includes labels.
Our method synthesizes new training data with a contrast similar to that of the test domain, through a process referred to as ``contrast harmonization" in MRI.
Our experiments show that combining one-shot adaptation data with harmonized training data outperformed the use of either one of the data sources alone.
Domain adaptation using only harmonized training data achieved comparable or even better performance compared to one-shot adaptation.
In addition, all adaptations only required light fine-tuning of 2 to 5 epochs for convergence.
\end{abstract}


\section{INTRODUCTION}
\label{sec:intro}  

Multiple sclerosis~(MS) is a central nervous system disorder characterized by inflammatory demyelination and axonal and neuronal degeneration \cite{haider2016topograpy}.
T2-weighted~(T2w) magnetic resonance imaging~(MRI) using the fluid-attenuated inversion recovery~(FLAIR) pulse sequence is routinely used for clinical diagnosis of MS lesions because it provides high lesion-to-brain contrast while simultaneously suppressing hyperintense cerebrospinal fluid~(CSF) signals, which can cause partial-volume artifacts in T2w images.~\cite{rovaris2000cortical}
Extensive manual editing is required for accurate delineation of MS lesions, though the task can be quite subjective.
Therefore, automatic detection and segmentation of MS lesions is desired for better efficiency and reproducibility.

State-of-the-art methods~\cite{zhang2019multiple, zhang2021all, valverde2019one, kamraoui2022deeplesionbrain, ackaouy2020unsupervised, zhao2021robust, aslani2020scanner, cerri2021contrast, billot2021joint, zhang2019rsanet, zhang2021efficient, zhang2021geometric} employ deep learning~(DL) to automate MS lesion segmentation using multi-contrast MRI scans, including FLAIR.
However, these algorithms frequently face challenges in achieving consistent performance across different MRI scanners and imaging protocols. 
This has led to increased research interest in domain adaptation techniques,
such as one-shot domain adaptation~\cite{valverde2019one}, spatially adaptive sub-networks~\cite{kamraoui2022deeplesionbrain}, domain-invariant latent feature learning~\cite{zhao2021robust, aslani2020scanner, ackaouy2020unsupervised}, contrast-adaptive generative modeling~\cite{cerri2021contrast}, and domain randomness via synthesis~\cite{billot2021joint}, to name just a few.

An alternative to domain adaptation is image harmonization, which reduces inter-site variation to aid downstream comparisons and analysis of images across imaging sites, scanners, and over time~\cite{fortin2017harmonization, fortin2018harmonization, hays2022spie}.
Synthesis-based multi-contrast MRI harmonization has shown remarkable progress in recent years~\cite{dewey2019deepharmony, dewey2020disentangled, zuo2021unsupervised, zuo2022disentangling}.
In this paper, we use HACA3~\cite{zuo2022haca3}, a new synthesis-based multi-site MRI harmonization approach, to enhance one-shot and even ``zero-shot'' domain adaptation performance for MS lesion segmentation.

\section{METHOD}
\label{sec:method}

\begin{figure}[!t]
	\centering
    \includegraphics[width=0.8\columnwidth,height=0.2916\columnwidth]{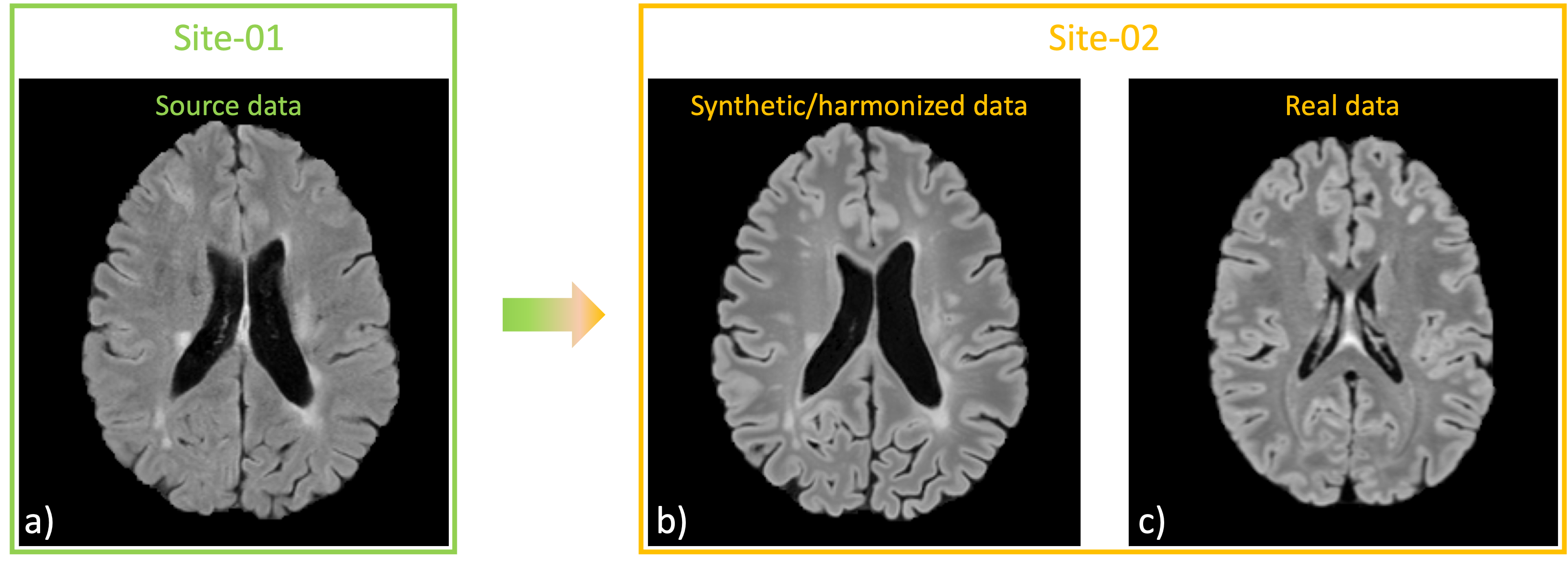}
	\caption{ 
	\textbf{Synthesis-based harmonization of one FLAIR axial slice.}   
	(a)~Source FLAIR slice on Site-01 (ISBI public training set).
	(b)~Harmonized FLAIR slice from Site-01 to Site-02 (in-house test set) using HACA3~\cite{zuo2022haca3}. 
	(c)~Real FLAIR slice on Site-02. 
	}
	\label{fig:figure1}
\end{figure}

\textbf{Dataset.}\quad 
We use the training data from the ISBI longitudinal dataset consisting of five people with MS~(PwMS)~\cite{carass2017longitudinal} and an in-house dataset including ten PwMS; both datasets have corresponding manual delineations.
We first pre-trained a segmentation network using four of the five training subjects from the ISBI dataset, with the remaining subject being used for validation.
We then applied the pre-trained segmentation network to our in-house dataset, which comes from a different domain than the training and validation data.

\noindent\textbf{Network.}\quad We implemented a modified 3D UNet following the design choices of nnUNet~\cite{isensee2021nnu}.
For the convolutional building block in the UNet, we chose the ``Conv+InstanceNorm+ReLU'' configuration with 2 blocks for each level of the encoding or decoding path of the UNet and 4 downsampling/upsampling operations. 
The numbers of channels in each convolutional block at all levels along the encoding path were 32, 64, 128, 256, and 512.
Three dimensional patches were cropped from skull-stripped\cite{isensee2019automated} and white matter intensity normalized~\cite{reinhold2019evaluating} T1-weighted~(T1w) and FLAIR images, and these patches were concatenated along the channel dimension to be used as input.
We generate the binary prediction of segmentation by thresholding the sigmoid of the output of UNet's last convolutional layer.

\noindent\textbf{Training.}\quad A batch size of two was used and the 3D patch size was set to $112\times112\times112$ to fully utilize GPU memory during backpropagation.
Heavy augmentations were employed on the fly, including random cropping, axis permutation, intensity shifts, as well as affine and elastic deformations.
The loss function was the mean of the Dice similarity coefficient~(DSC) and binary cross-entropy.
The Adam optimizer was used with an initial learning rate of $10^{-4}$ for 100 epochs, where each epoch involved the application of eight random augmentations to every training subject.

\noindent\textbf{Harmonization-based domain adaptation.}\quad For domain adaptation after training, all T1w and FLAIR images from the ISBI training dataset were transformed to match the contrast of our in-house test dataset using the synthesis-based multi-site harmonization of HACA3~\cite{zuo2022haca3}. HACA3 was trained on diverse MR datasets acquired from 21 sites, which included white matter lesion multi-contrast images with varying field strengths, scanner platforms, and acquisition protocols.
An example of such image harmonization is shown in Fig.~\ref{fig:figure1}, where the FLAIR ``source data'' from ``Site-01'' in the ISBI training set was harmonized to match the contrast of ``Site-02" of our in-house test set.
Consistent gray and white matter contrast was observed between ``synthetic/harmonized data'' and ``real data'' for ``Site-02". 

After harmonization, three domain adaptation strategies were evaluated:
\bq{\makebox[3.7cm][l]{\textbf{One-Shot Strategy}} Fine-tune~(FT) the pre-trained network with only one of the ten subjects in the test domain and evaluate on the remaining nine test subjects.}
\bq{\makebox[3.7cm][l]{\textbf{Zero-Shot Strategy}} FT with only the harmonized ISBI data and evaluate on the ten test subjects.}
\bq{\makebox[8.0cm][l]{\textbf{Harmonization-enriched One-Shot Strategy}} FT with a combination of all harmonized ISBI training data and one of the test domain subjects, and evaluate on the remaining nine test subjects.}

For the One-Shot and Harmonization-enriched One-Shot strategies, ten-fold cross-validation~(CV) was employed to evaluate on all ten test subjects from the in-house data, where each of the ten subjects in the test domain was included in the training set for one fold.
All FTs were conducted for 20 epochs, and the models were tested after every epoch. 
For comparison, two-fold CV FT with 4/1/5 training/validation/test subjects per fold was also employed for 100 epochs.
This CV served as a ``normal'' performance estimation, assuming enough labeled data in new domains was provided for network adaptation.

\noindent\textbf{Evaluation metrics.}\quad The DSC, lesion-wise F1 score~(L-F1), and Pearson’s correlation coefficient of the lesion volumes between the ground truth and the prediction~(VC) were utilized as the segmentation performance evaluation metrics in the experiment.

\section{RESULTS}
\label{sec:experiments}

\begin{figure}[!t]
	\centering
    \includegraphics[width=\columnwidth,height=0.1777\columnwidth]{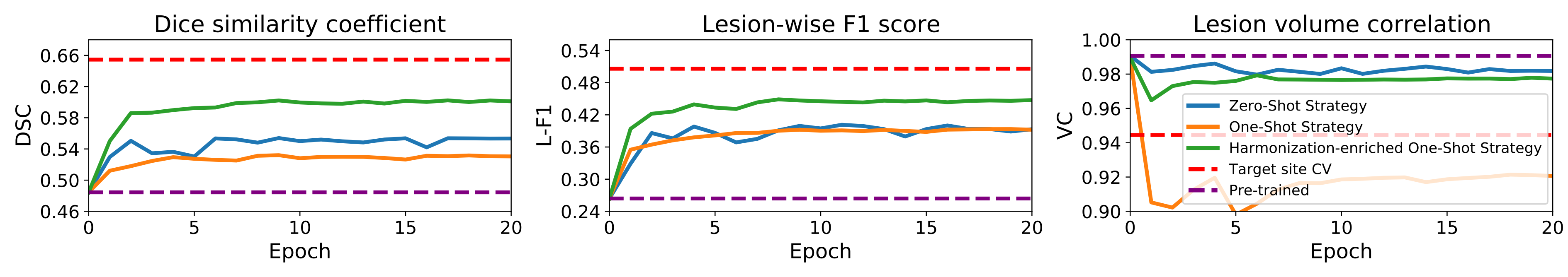}
	\caption{ 
	\textbf{Quantitative metrics assessed after each epoch of domain adaptation.}
        The performance of all three strategies converges after only two to five epochs, with noticeable improvement from pre-trained results~(dashed purple lines) in DSC and L-F1.
        Harmonization-enriched One-Shot adaptation data~(solid green lines) consistently outperformed the use of either the One-Shot~(solid orange lines) or Zero-Shot~(solid blue lines) strategies in terms of DSC and L-F1. Notably, the Harmonization-enriched One-Shot strategy achieved a DSC score of above 0.6 after convergence, which is close to inter-rater consistency~\cite{carass2020sr}.
	}
	\label{fig:figure2}
\end{figure}

\noindent\textbf{Quantitative comparison.}\quad Figure~\ref{fig:figure2} shows DSC, L-F1, and VC scores of the three domain adaptation strategies after each FT epoch. 
First, the performance of all three strategies converged within just 2 to 5 epochs, exhibiting noticeable improvement over the pre-trained results (dashed purple lines) in terms of DSC and L-F1.
However, a noticeable degradation in VC was observed for normal CV (dashed red line) and the One-Shot strategy~(solid orange line), which was not observed for the Zero-Shot~ (solid blue line) and Harmonization-enriched One-Shot~(solid green line) strategies.
Second, the Harmonization-enriched One-Shot strategy consistently outperformed the other two strategies in terms of DSC and L-F1, and performed similarly well with the Zero-Shot strategy for VC.
Notably, the Harmonization-enriched One-Shot strategy achieved a DSC score of above 0.6 after convergence, approaching inter-rater consistency~\cite{carass2020sr}.
Third, Zero-Shot outperformed One-Shot strategy in terms of DSC and VC, while these two strategies performed similarly for L-F1.

\begin{figure}[!t]
	\centering
    \includegraphics[width=\columnwidth,height=0.4724\columnwidth]{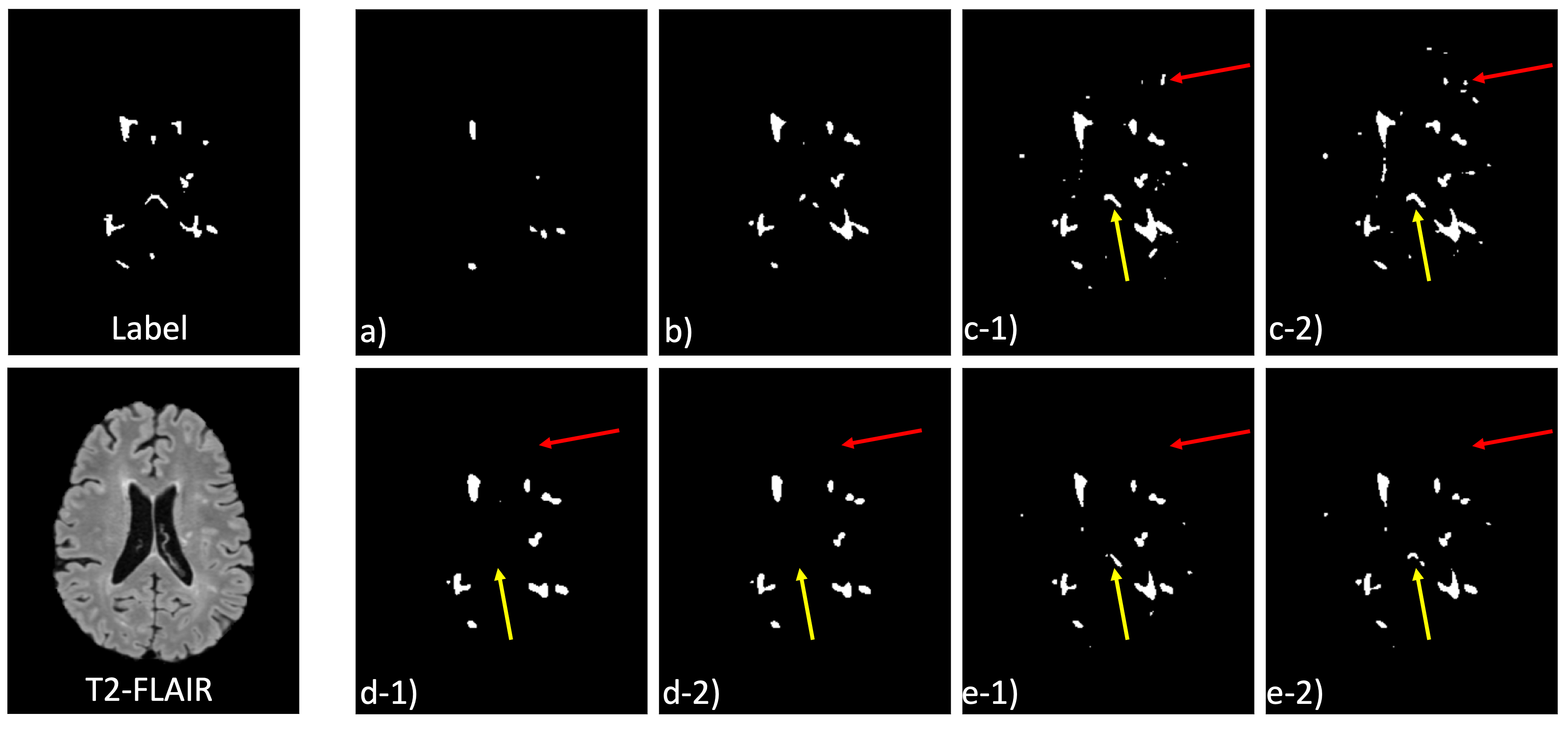}
	\caption{ 
	\textbf{Segmentation predictions with ground-truth label and axial FLAIR slice.}  
	(a)~Pre-trained prediction.
	(b)~Target site CV. 
	(c)~One-shot adaptation after 2~(c-1) and 20 epochs~(c-2).
        (d)~Our Zero-Shot adaptation after 2~(d-1) and 20 epochs~(d-2).
        (e) Harmonization-enriched One-Shot adaptation after 2~(e-1) and 20 epochs~(e-2).
        False positive lesions (red arrows) in~(c) were avoided in the Zero-Shot~(d) and Harmonization-enriched One-Shot~(e) strategies. False negative lesions are denoted with the yellow arrows.
	}
	\label{fig:figure3}
\end{figure}

\noindent\textbf{Qualitative comparison.}\quad Figure~\ref{fig:figure3} shows segmentation predictions with the corresponding ground-truth label and axial FLAIR slice~(left column).
The pre-trained prediction (Fig.~\ref{fig:figure3}a) exhibited false negative predictions throughout the entire image, which was addressed by the target site CV (Fig.~\ref{fig:figure3}b).
False positive lesions (indicated by red arrows) were observed in the One-Shot strategy after 2 (Fig.~\ref{fig:figure3}c-1) and 20 (Fig.~\ref{fig:figure3}c-2) fine-tuning epochs, which were not observed in the Zero-Shot strategy~(Fig.~\ref{fig:figure3}d) or the Harmonization-enriched One-Shot strategy~(Fig.~\ref{fig:figure3}e) after 2 epochs (Fig.~\ref{fig:figure3}d-1 and Fig.~\ref{fig:figure3}e-1, respectively) or 20-epochs (Fig.~\ref{fig:figure3}d-2 and Fig.~\ref{fig:figure3}e-2) of fine-tuning.
False negative lesions (indicated by yellow arrows) missed by the Zero-Shot strategy (Fig.~\ref{fig:figure3}d) were still captured by the One-Shot and Harmonization-enriched One-Shot strategies (Fig.~\ref{fig:figure3}c and Fig.~\ref{fig:figure3}e, respectively).
%
%
No significant differences were observed between the 2 epoch (Fig.~\ref{fig:figure3}d-1 or Fig.~\ref{fig:figure3}e-1) and 20 epoch (Fig.~\ref{fig:figure3}d-2 or Fig.~\ref{fig:figure3}e-2) fine-tuning results of the Zero-Shot or Harmonization-enriched One-Shot strategies.

\section{New work to be presented}
\label{sec:breakthrough}
We will present how to use synthesis-based harmonization to boost one-shot domain adaptation performance for MS lesion segmentation. This work has not been submitted or presented elsewhere before.

\section{DISCUSSION AND CONCLUSION}
\label{sec:discussion}
Domain adaptation through network fine-tuning has been successfully applied to different imaging problems in MRI, including under-sampled k-space reconstruction~\cite{zhang2020fidelity}, biophysical inversion~\cite{zhang2021hybrid} with uncertainty quantification~\cite{zhang2020bayesian}, and contrast translation~\cite{he2021autoencoder}.
In this work, we demonstrate the feasibility of leveraging synthesis-based MRI harmonization to enhance domain adaptation performance in MS lesion segmentation.
Our experiments demonstrate that our Zero-Shot domain adaptation, utilizing solely public data synthesized to the target contrast, yields comparable or superior performance than a One-Shot strategy on the target domain.
More notably, the combination of One-Shot and Zero-Shot adaptation, which we coin as Harmonization-enriched One-Shot domain adaptation, achieved DSC results approaching inter-rater performance.
Additionally, only light fine-tuning of between 2 and 5 epochs was enough for an adequate adaptation of the pre-trained network.

\acknowledgments 
This material is partially supported by the National Science Foundation Graduate Research Fellowship under Grant No. DGE-1746891 (Remedios).
This work also received support from National Multiple Sclerosis Society RG-1907-34570 (Pham), CDMRP W81XWH2010912 (Prince), and the Department of Defense in the Center for Neuroscience and Regenerative Medicine.
The opinions and assertions expressed herein are those of the authors and do not reflect the official policy or position of the Uniformed Services University of the Health Sciences or the Department of Defense.

\bibliography{report} 
\bibliographystyle{spiebib} 

\end{document}